\documentclass[a4paper]{jpconf}
\usepackage{graphicx}
\usepackage{epstopdf}
\newcommand{\sNN}[1]{$\sqrt{s_{NN}} = #1$ GeV}
\newcommand{\Lam}{$\Lambda $ }
\newcommand{\ALam}{$\bar{\Lambda} $ }
\newcommand{\Kaon}{$K^{0}_{S} $ }

\begin{document}
\title[Strangeness production in jets from p+p \sNN{200} collisions]{Strangeness production in jets from p+p $\sqrt{s}  = 200$ GeV collisions}
\author{Anthony R. Timmins for the STAR Collaboration}
\address{Department of Physics and Astronomy, Wayne State University, 666 W. Hancock, Detroit, MI 48201, USA}
\ead{tone421@rcf.rhic.bnl.gov}

\begin{abstract}

Measurements of strangeness production in jets help illuminate the QCD mechanisms in fragmentation. Furthermore, they provide a crucial baseline for heavy-ion studies where modifications in jet chemistry have recently been predicted. We present new results on strange particle production in jets from p+p $\sqrt{s}$ = 200 GeV collisions measured by the STAR experiment. The momentum distributions of the \Lam, \ALam and \Kaon particles are obtained using various jet finding algorithms, and then compared to various models. Strange particle ratios in jets are obtained and compared to values obtained from the inclusive spectra. Finally, we show jets tagged with leading strange baryons and mesons, in order to investigate whether gluon or quark jets can be isolated in this way.

\end{abstract}

\section{Introduction}

We report measurements of strange particle momentum distributions in jets from p+p $\sqrt{s}  = 200$ collisions measured by the STAR experiment. These results provide additional tests to QCD inspired models of fragmentation such as PYTHIA \cite{PYTHIA} or the Modified Leading Log Approximation (MLLA) \cite{MLLA1,MLLA2,MLLA4,MLLA5}, beyond the usual comparisons to charged hadron momentum distributions. Furthermore, predictions for the modification of jet chemistry in heavy-ion collisions relative to p+p have been made \cite{SapWeid}. These predictions show the $K/\pi$ and $p/\pi$ ratios in jets are expected to be higher in the presence of a medium. It is therefore essential we test theoretical descriptions of identified particle momentum distributions in p+p jets, if predictions in the modification of heavy-ion jet chemistry are to be made. Finally, since jets are from hard processes, comparisons of particles yields in jets to inclusive spectrum measurements may help elucidate the contributions from hard processes to those measurements as a function of $p_{T}$.

\begin{figure}[b]
\begin{center}
\includegraphics[width = 1\textwidth]{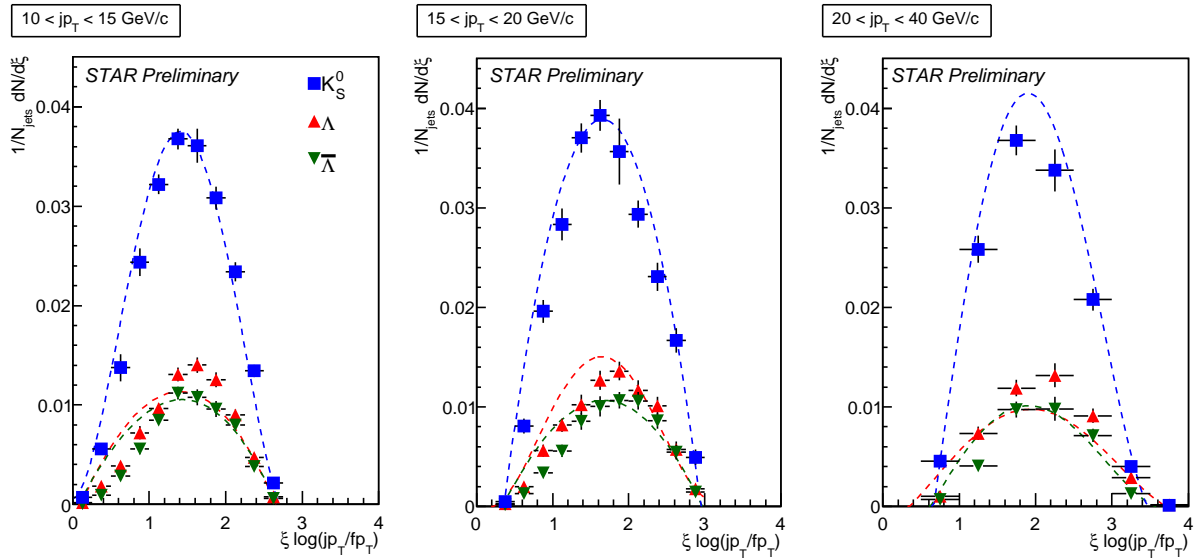}
\end{center}
\caption{Data points show strange particle momentum distributions for three reconstructed jet $p_{T}$ ranges in p+p \sNN{200} collisions. The strange particle yields are not corrected for tracking efficiency. The symbols $jp_{T}$ and $fp_{T}$ are the reconstructed jet and fragment $p_{T}$ respectively. The curves show polynomial fits to PYTHIA 6.4 predictions. These predictions are obtained from PYTHIA events ran through the STAR detector simulator. The same applies to all other figures.} 
\label{fig1} 
 \end{figure}

\section{Analysis}

The data presented are from $\sim$ 8 million jet-patch triggered p+p \sNN{200} events recorded in 2006 by the STAR experiment. The jet-patch trigger is designed to select events with jets, and thus requires energies above 8 GeV to be deposited in an area $\Delta \eta \times \Delta \phi = 1 \times 1$ within the electromagnetic calorimeter \cite{STARCal}. The Time Projection Chamber (TPC) is used to detect the charged particles \cite{TPC}, while the barrel electromagnetic calorimeter, BEMC, is used for neutral particles. Three jet finders are employed from the fast-jet package \cite{fastjet}: $k_{t}$, anti-$k_{t}$ and SIS. We select recoil jets with respect to the triggered jets to measure fragmentation functions. This is because the triggered jets have a bias towards the neutral jet component. The jets are found within a resolution parameter, R $=0.4$ and a jet axis $\eta$ cut $< 0.6$ to ensure the jet area lies within the TPC/BEMC acceptance. The resolution parameter can be thought of as an effective jet radius in the $\eta,\phi$ plane. The minimum $p_{T}$ cut for a charged particle is 200 MeV/c, while the minimum calorimeter tower energy is 200 MeV. Towers which match charged particles are removed from the jet to avoid over counting of electrons and Minimum Ionising Particles (MIPs). 
No corrections are applied to the jet energy scale which means the reconstructed jet $p_T$ is likely to be less than the actual jet $p_{T}$ due to TPC/BEMC detection inefficiencies, and missing neutral energy principally from undetected $n$ and $K^{0}_{L}$ particles. Studies are underway to determine a correction for this. Finally, in order to measure the momentum distributions of the $\Lambda$, $\bar{\Lambda}$, and $K^{0}_{S}$ particles, V0s are identified within the jet area from TPC tracks. The invariant mass distributions are calculated to extract the yields of  these particles via their dominant decay channels: $\Lambda \rightarrow p+\pi^{-}$, $\bar{\Lambda} \rightarrow \bar{p}+\pi^{+}$, $K^{0}_{S} \rightarrow \pi^{+}+\pi^{-}$. In conjunction with a minimum $p_T$ cut of 1 GeV/c, a series of topological cuts are placed on the V0s to minimise the relative background contribution, and ensure the signal to background ratios are approximately the same for each particle. The residual background after all cuts is not yet removed from the fragment yields.

\section{Results}

Strange particle momentum distributions as a function of the reconstructed jet $p_T$ are shown in figure \ref{fig1}. The momentum distribution is expressed in $\xi=log(jp_{T}/fp_{T})$, where $jp_{T}$ and $fp_{T}$ are the reconstructed jet and fragment $p_{T}$ respectively.  For a particular particle species, the integral of the $\xi$ function gives its mean multiplicity per jet. The uncertainties are statistical added in quadrature to the small differences obtained by comparing the three different jet finders. A data point is the mid point value between the lowest and highest values the three jet finders give. The measured V0 yields are not yet corrected for acceptance and TPC tracking inefficiencies. The PYTHIA predictions are obtained by processing PYTHIA events with STAR's detector simulation, and running the track, jet, and V0 reconstruction software on the simulated output in the same way as the real data. This ensures the PYTHIA events are subject to the same acceptance restrictions/detector inefficiencies as the real data.  All PYTHIA predictions shown in these proceedings are from version 6.4 with Tune A. We find that PYTHIA gives a reasonable description of the \Kaon data. JETSET (the jet production scheme in PYTHIA) has also been shown to describe $K^{\pm}$ fragmentation functions in $e^{+}+e^{-}$ collisions well, where the jet energies range $5 < E_{jet} < 46$ GeV \cite{FAnulli}. Furthermore, PYTHIA has been shown to describe charged hadron momentum distributions in p+p collisions at the same center of mass energy  \cite{HelenQM}. On the other hand, we note that the PYTHIA description of \Lam and \ALam  momentum distributions are less satisfactory. Although the predictions appear to predict the correct yield over all $\xi$, they tend to over predict the yields at low $\xi$ and under predict the yields at intermediate $\xi$. Whether this 
is due to a mismatch in the reconstruction of PYTHIA events and real data, or due to physics, is under investigation.

\begin{figure}[h]
\begin{center}
\includegraphics[width = 1.0\textwidth]{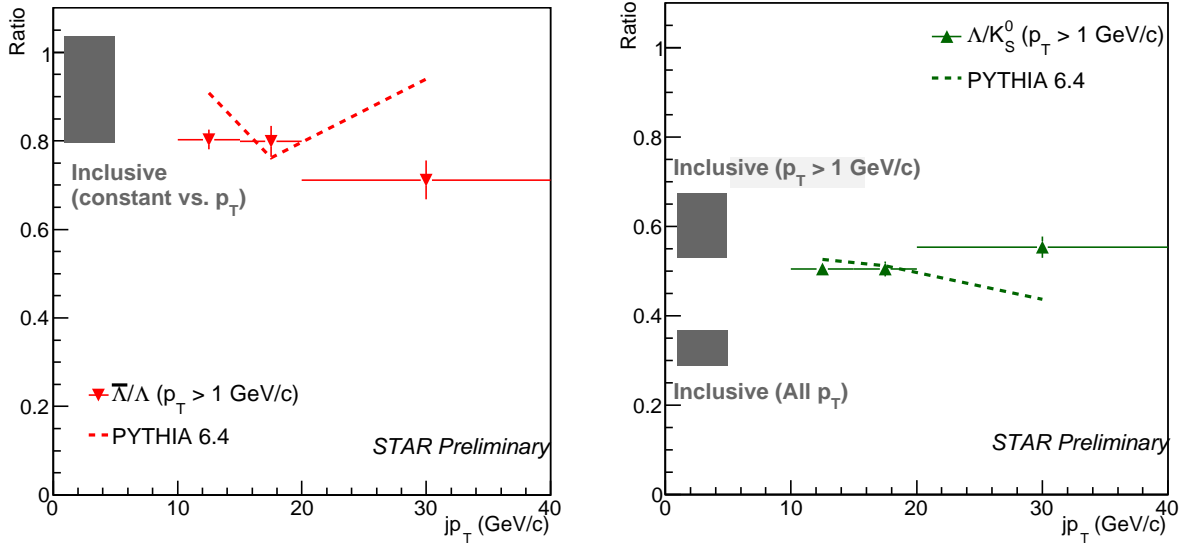}
\end{center}
\caption{ $\bar{\Lambda}/\Lambda$ and $\Lambda/K^{0}_{S}$ ratios in jets for three reconstructed jet $p_{T}$ ranges in p+p \sNN{200} collisions. In each case, the minimum particle $p_{T}$ is 1 GeV/c and ratios are corrected for TPC efficiency and acceptance. The uncertainties are statistical added in quadrature to the small differences obtained by comparing the three different jet finders. Inclusive refers to values obtained from inclusive spectra measurements \cite{STARpp}. The vertical size of shaded boxes reflects statistical and systematic uncertainties added in quadrature.}
\label{fig2} 
 \end{figure}

In figure \ref{fig2}, we show ratios of integrated yields $\bar{\Lambda}/\Lambda$ and $\Lambda/K^{0}_{S}$ for $p_{T} > 1$ GeV/c as a function of reconstructed jet $p_T$. The analysis shows the $\bar{\Lambda}/\Lambda$ ratio is below 1 showing the baryon asymmetry in the colliding system (p+p) is transferred to jets over all jet energies measured. We also find the $\bar{\Lambda}/\Lambda$ ratio is consistent with the value obtained from inclusive spectra measurements \cite{STARpp}. PYTHIA also seems to reproduce the magnitude of the ratio. The jaggedness of the predictions is due statistical fluctuations which are $\sim 10\%$ and not shown. Regarding the $\Lambda/K^{0}_{S}$ ratios in jets, these are higher than values from inclusive spectrum measurements over all $p_{T}$. The inclusive spectrum measurement will be dominated by production below 1 GeV/c. When we compare the ratios in jets to the inclusive spectrum measurement above 1 GeV/c, we find they are generally consistent. This may mean that $\Lambda$ and $K^{0}_{S}$ spectrum measurements with $p_{T} > 1$ GeV/c have a dominant contribution from hard processes i.e. jet production. However, measurements of the same ratio from the underlying event are needed to further qualify this observation. We also find PYTHIA again reproduces the magnitude of the ratio.

\begin{figure}[h]
\begin{center}
\includegraphics[width = 0.6\textwidth]{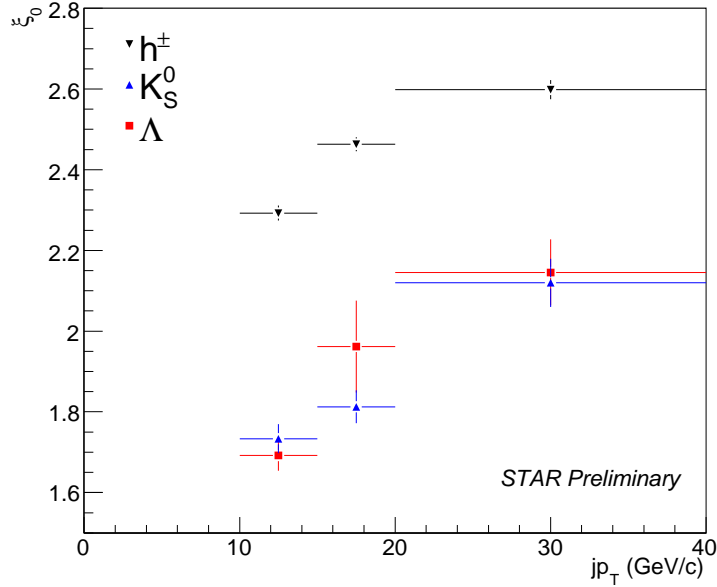}
\end{center}
\caption{Peak position of $\xi$ ($\xi_{0}$) for charged hadrons, \Lam and \Kaon particles for three reconstructed jet $p_{T}$ ranges in p+p \sNN{200} collisions. The uncertainties are statistical added in quadrature to the small differences obtained by comparing the three different jet finders.  }
\label{fig3} 
 \end{figure}

In figure \ref{fig3}, we show the position of the maximum of the $\xi$ distribution, $\xi_{0}$, for the various particles as a function of the reconstructed jet $p_{T}$. Since the $p_{T}$ dependent tracking inefficiency/acceptance will distort the shape of the $\xi$ distribution, we correct the distributions using embedding which determines the V0 reconstruction efficiency. A Gaussian is fitted in the peak region to extract $\xi_{0}$. The fit range is chosen to avoid the region in $\xi$ affected by the minimum $p_T$ cut on the fragments. In the MLLA scheme, the position of the maximum can be described by the following relation \cite{MLLA5};
\begin{equation}
\xi_{0} = Y + \sqrt{cY}-c
\label{equ:MLLA1}
\end{equation}
where $c$ is a constant and $Y$ is:
\begin{equation}
Y=log \left( \frac{E_{jet}sin(\theta_{0})}{Q_{0}} \right)
\label{equ:MLLA1}
\end{equation}
where $\theta_{0}$ is the jet opening angle, and $Q_{0}$ is the effective transverse momentum (relative to the jet axis) where a parton ceases to branch in the parton shower. MLLA predictions for momentum distributions of hadrons with mass $M_{0}$ typically assume $Q_{0} \simeq M_{0}$ \cite{SapWeid}. Equation \ref {equ:MLLA1} then implies a scaling behaviour of $\xi_{0}$ with the mass of the hadron i.e.  $\xi_{0}$ should decrease for the higher mass hadrons. The value of $\xi_{0}$ should also increase with jet energy for a given hadron which is observed in figure \ref{fig3} for all particles. However, we find that  $\xi_{0}(K^{0}_{S}) \sim \xi_{0}(\Lambda)$ which is contrary to the expected scaling. On the other hand, $\xi_{0}(h^{\pm}) > \xi_{0}(K^{0}_{S})$ and $\xi_{0}(\Lambda)$ which is consistent with such a scaling assuming pions dominate particle production in jets. A similar observation was made for $\pi,K,p$ fragmentation functions for similar jet energies in $e^{+}+e^{-}$ collisions \cite{FAnulli1}. Finally, we note in interest that if we extract the mode of the transverse rapidity, $y_{T}$, distribution (motivated by this study \cite{TrainorFF}), we find a mass scaling in $y_{T}$ space i.e $y_{T,0}(h^{\pm}) >  y_{T,0}(K^{0}_{S})) > y_{T,0} (\Lambda)$ for a given reconstructed jet $p_{T}$ bin. For a given particle species, we also find $y_{T,0}$ increases with jet energy.

\begin{figure}[h]
\begin{center}
\includegraphics[width = 1\textwidth]{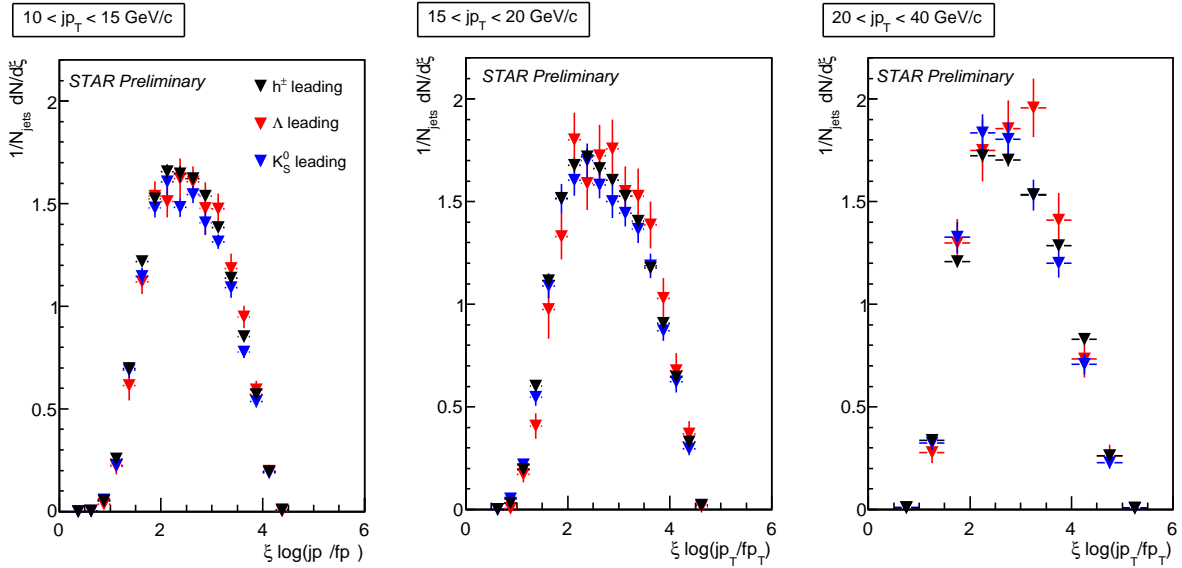}
\end{center}
\caption{Momentum distributions of non-leading charged hadrons for three reconstructed jet $p_{T}$ ranges in p+p \sNN{200} collisions. The black points show data for jets where the leading particle is a charged hadron, red corresponds to jets with leading \Lam particles, and the blue points correspond to jets with leading \Kaon particles.  The charged hadron yields are not corrected for tracking efficiency. $jp_{T}$ and $fp_{T}$ are the reconstructed jet and fragment $p_{T}$ respectively.  The uncertainties are statistical added in quadrature to the small differences obtained by comparing the three different jet finders.  } 
\label{fig4} 
 \end{figure}

In figure \ref{fig4} we show the momentum distributions of \emph{non-leading} charged hadrons for the cases where the leading particle is a charged hadron, $\Lambda$, or $K^{0}_{S}$. We want to investigate whether tagging jets according to the species of the leading hadron, preferentially selects on gluon or quark jets. Both measured data \cite{DELPHJets} and theoretical predictions show hadron multiplicities are larger in gluon jets compared to quark jets: MLLA gives the ratio 9/4 for gluon/quark jet multiplicities. Thus, if tagging jets with particle A preferentially selected gluon jets, and tagging jets with particle B preferentially selected quark jets, we would expect the jets associated with particle A to have more non-leading charged hadrons compared to jets associated with particle B. However, in figure \ref{fig4} we observe at given jet energy, the charged hadron multiplicities are the same for each of the tagged jets. This might be expected for the highest jet $p_T$ bin where jet production is expected to be dominated by hard scattering of valance quarks, however at lower jet energies there should be mixture of quark and gluon jets \cite{GluonQuark}. Therefore, our observation suggests quark/gluon jets cannot be tagged in this way. Further studies are underway to confirm this.

\section{Summary}

In summary, we have shown measurements of strange particle momentum distributions. We have found that PYTHIA describes the $K^{0}_{S}$ momentum distributions well, however we observe some deviations for the $\Lambda$ and $\bar{\Lambda}$ data. We have found that $\bar{\Lambda}/\Lambda$ ratios ($p_{T} > 1$ GeV/c) in jets are consistent with values obtained from inclusive spectra. ${\Lambda}/K^{0}_{S}$ ratios  ($p_{T} > 1$ GeV/c) in jets were found to be similar to values obtained from inclusive spectra in the same $p_T$ range. This may suggest that ${\Lambda}$ and $K^{0}_{S}$ production above 1 GeV/c is dominated by hard processes in p+p \sNN{200} collisions. We investigated a mass scaling of the peak position of $\xi$ (inferred from MLLA), and found although $\xi_{0}(h^{\pm}) > \xi_{0}(K^{0}_{S})$ and $\xi_{0}(\Lambda)$, $\xi_{0}(K^{0}_{S}) \sim \xi_{0}(\Lambda)$. Finally, we have found tagging jets with leading strange baryons/mesons may not preferentially select on quark/gluon jets.


\section*{References}
\begin{thebibliography}{10}
%

\bibitem{PYTHIA} Sjostrand T, Mrenna S, and Skands P 2006  \emph{JHEP}  {\bf 0605} 026 
%

\bibitem{MLLA1} Dokshitzer Y, Fadin V, and Khoze V 1982  \emph{Phys. Lett. B} {\bf 115 } 242; Bassetto A, Ciafaloni M, Marchesini G, and Mueller A 1982 \emph{Nucl. Phys. B} {\bf 207}
\bibitem{MLLA2}  Mueller A 1983 \emph{Nucl. Phys. B} {\bf 213} 85  
\bibitem{MLLA4}  E Malaza  E and Webber B (1984) Phys. Lett. B  {\bf 149} 501
\bibitem{MLLA5}  Dokshitzer Y, Mueller A , Khoze V, and Troyan S, \emph{Basics of Perturbative QCD (Editions Frontieres, Paris, 1991)}
%
\bibitem{SapWeid} Sapeta S and Wiedemann U 2008 \emph{Eur. Phys. J. C} {\bf 55} 293
%
\bibitem{STARCal} Abelev B \emph{et al.} (STAR Collaboration), 2008 \emph{Phys. Rev. Lett.} {\bf 100} 232003
\bibitem{TPC} Ackermann K H \emph{et al.},  2003 \emph{Nucl. Inst. and Meth. A} {\bf 499} 659
\bibitem{fastjet} Salam G and Soyez G 2007 \emph{JHEP} {\bf 0705} 086
\bibitem{FAnulli} Anulli F (BABAR Collaboration) \emph{hep-ex/0406017v1}
\bibitem{HelenQM} Caines H  (STAR Collaboration) 2009 \emph{Nucl. Phys. A} {\bf 830} 263c-266c
\bibitem{STARpp} Abelev B \emph{et al.} (STAR Collaboration) 2007 \emph{Phys. Rev.} C {\bf 75} 064901
\bibitem{FAnulli1} Albino S \emph{et al.} \emph{arXiv:0804.2021v1}
\bibitem{TrainorFF} Trainor T and Kettler D 2006 \emph{Phys. Rev. D} {\bf 74} 034012
\bibitem{DELPHJets} Abreu P (DELPHI Collaboration) 2000 \emph{Eur. Phys. J. C} {\bf 13} 573
\bibitem{GluonQuark} De Florian D and Vogelsang W 2007 \emph{Phys. Rev. D} {\bf 76 } 074031

\end{thebibliography}
\end{document}